\documentclass[%
 reprint,
superscriptaddress,
 amsmath,amssymb,
 aps,
pra,
]{revtex4-2}

\usepackage{graphicx}
\usepackage{dcolumn}
\usepackage{bm}
\usepackage[unicode=true,
 bookmarks=true,bookmarksnumbered=false,bookmarksopen=false,
 breaklinks=false,pdfborder={0 0 1},backref=false,colorlinks=true]{hyperref}

\usepackage{physics, color}
\hypersetup{linkcolor=magenta, urlcolor=blue, citecolor=blue, pdfstartview={FitH}}

\usepackage{soul}

\begin{document}

\preprint{APS/123-QED}

\title{Practical Quantum Reservoir Computing in Rydberg Atom Arrays}
\author{Dong-Sheng Liu}
\affiliation{Laboratory of Quantum Information, University of Science and Technology of China, Hefei 230026, China}
\affiliation{Center For Excellence in Quantum Information and Quantum Physics, University of Science and Technology of China, Hefei 230026, China}
\affiliation{Anhui Province Key Laboratory of Quantum Network, University of Science and Technology of China, Hefei 230026, China.}
\affiliation{Hefei National Laboratory, University of Science and Technology of China, Hefei 230088, China.}

\author{Qing-Xuan Jie}
\affiliation{Laboratory of Quantum Information, University of Science and Technology of China, Hefei 230026, China}
\affiliation{Center For Excellence in Quantum Information and Quantum Physics, University of Science and Technology of China, Hefei 230026, China}
\affiliation{Anhui Province Key Laboratory of Quantum Network, University of Science and Technology of China, Hefei 230026, China.}

\author{Chang-Ling Zou}
\email{clzou321@ustc.edu.cn}
\affiliation{Laboratory of Quantum Information, University of Science and Technology of China, Hefei 230026, China}
\affiliation{Center For Excellence in Quantum Information and Quantum Physics, University of Science and Technology of China, Hefei 230026, China}
\affiliation{Anhui Province Key Laboratory of Quantum Network, University of Science and Technology of China, Hefei 230026, China.}
\affiliation{Hefei National Laboratory, University of Science and Technology of China, Hefei 230088, China.}

\author{Xi-Feng Ren}
\email{renxf@ustc.edu.cn}
\affiliation{Laboratory of Quantum Information, University of Science and Technology of China, Hefei 230026, China}
\affiliation{Center For Excellence in Quantum Information and Quantum Physics, University of Science and Technology of China, Hefei 230026, China}
\affiliation{Anhui Province Key Laboratory of Quantum Network, University of Science and Technology of China, Hefei 230026, China.}
\affiliation{Hefei National Laboratory, University of Science and Technology of China, Hefei 230088, China.}

\author{Guang-Can Guo}
\affiliation{Laboratory of Quantum Information, University of Science and Technology of China, Hefei 230026, China}
\affiliation{Center For Excellence in Quantum Information and Quantum Physics, University of Science and Technology of China, Hefei 230026, China}
\affiliation{Anhui Province Key Laboratory of Quantum Network, University of Science and Technology of China, Hefei 230026, China.}
\affiliation{Hefei National Laboratory, University of Science and Technology of China, Hefei 230088, China.}

\date{\today}

\begin{abstract}
Quantum reservoir computing (QRC) is a promising quantum machine learning framework for near-term quantum platforms, yet the performance of different QRC architectures under realistic constraints remains largely unexplored. Here, we provide a comparative numerical study of single-step-QRC (SS-QRC) and multi-step-QRC (MS-QRC) architectures implemented on a Rydberg atom array. We demonstrate that while MS-QRC performance is highly sensitive to the underlying dynamical phase of matter and decoherence, SS-QRC exhibits greater robustness. Using the randomized measurement toolbox to mitigate measurement overhead, we reveal that sampling noise undermines the convergence property required for MS-QRC. This leads to a significant reduction in the information processing capacity (IPC) of MS-QRC, deteriorating its performance on nonlinear time-series benchmarks. In contrast, SS-QRC maintains high IPC and accuracy across both temporal and non-temporal tasks. Our results suggest SS-QRC as a preferred candidate for near-term practical applications due to its resilience to system configurations and statistical noise.
\end{abstract}

\maketitle


\section{Introduction}
Quantum machine learning offers a promising pathway toward accelerating data analysis by leveraging quantum superposition and entanglement~\cite{Schuld2021a, Biamonte2017}. However, many early quantum machine learning algorithms rely on fault-tolerant quantum computers and quantum random-access memory, which are unavailable in the near term~\cite{Pastorello2023}. Consequently, realizing a practical quantum advantage requires a transition toward algorithms compatible with noisy intermediate-scale quantum (NISQ) devices~\cite{Bharti2022}.

Variational quantum algorithms, which utilize parameterized quantum circuits for information processing, have emerged as leading candidates for demonstrating near-term quantum advantages~\cite{Cerezo2021}. The power of these models typically scales with the expressivity of the parameterized quantum circuits, i.e., the capability to explore a vast region of the Hilbert space. However, a fundamental trade-off exists between expressivity and trainability: it has been proven that highly expressive ansatzes are significantly more susceptible to the barren plateau phenomenon, where cost function gradients vanish exponentially with the system size~\cite{McClean2018, Holmes2022}. While various initialization strategies have been proposed to mitigate barren plateaus, these are often ineffective against noise-induced barren plateaus~\cite{Wang2021}, necessitating complex error mitigation schemes that increase computational overhead~\cite{Ge2022}.

Quantum reservoir computing (QRC)~\cite{Mujal2021, Ghosh2021b} offers a promising alternative for near-term applications, efficiently circumventing the aforementioned training challenges. This framework is an extension of the paradigms of classical reservoir computing~\cite{Lukosevicius2009} and extreme learning machines~\cite{Huang2011} into the quantum domain, which uses a quantum system as the reservoir for information processing. By fixing the parameters of the quantum system and using its complex dynamics as a high-dimensional feature map, QRC requires training only a classical linear readout layer, significantly relaxing the control and operational requirements on quantum systems.

\begin{figure*}
    \centering
    \includegraphics[width=0.65\linewidth]{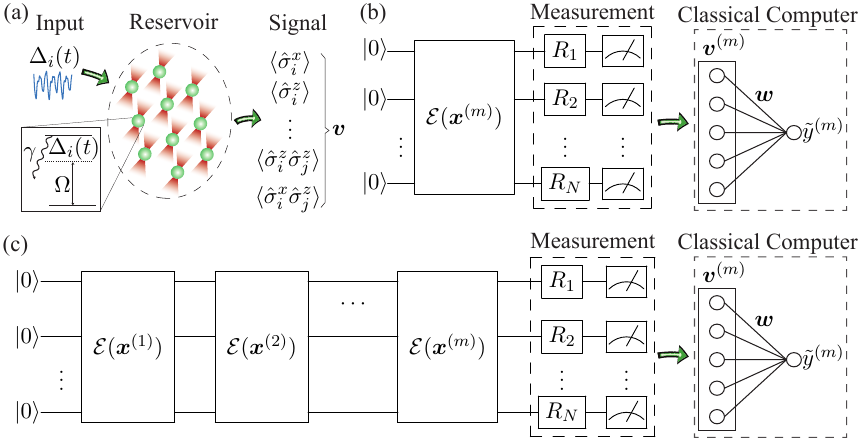}
    \caption{Schematic of quantum reservoir computing (QRC) with neutral atom arrays. (a) The reservoir consists of neutral atoms trapped in optical tweezers. Global Rydberg excitation drives atom-atom interactions, nonlinearly mapping inputs encoded in local detunings $\Delta_i(t)$ into an exponentially large Hilbert space. Outputs are extracted by measuring single- and two-site Pauli operators. (b) The single-step-QRC (SS-QRC) architecture. For each input $\boldsymbol{x}^{(m)}$, the reservoir evolves via a single CPTP map $\mathcal{E}(\boldsymbol{x}^{(m)})$. The measured signals $\boldsymbol{v}^{(m)}$ are processed by a linear layer with weights $\boldsymbol{w}$ trained via ridge regression. (c) The multi-step-QRC (MS-QRC) architecture. The reservoir processes a time-series sequence $\boldsymbol{x}^{(1)},\dots,\boldsymbol{x}^{(m)}$ through sequential evolution to capture temporal dependencies, followed by measurement and linear post-processing.}
    \label{fig1}
\end{figure*}

Despite the great potential of QRC, several critical factors should be taken into account to make practical use of this framework. First, the dynamical phase of matter in the quantum system has been shown to influence QRC performance~\cite{Martinez-Pena2021, Xia2022}. Localized phases typically favor linear memory, whereas ergodic phases enhance nonlinear processing capability. Second, decoherence is an inherent feature of NISQ hardware. While some level of decoherence noise can facilitate the fading memory property required for reservoir computing~\cite{Kubota2023a, Domingo2023, Fry2023}, excessive decoherence causes rapid information dissipation. Third, the sampling noise arising from a finite number of measurement shots $N_s$ introduces statistical fluctuations that can degrade the reservoir's performance~\cite{Hu2023}. Finally, while various QRC architectures have been proposed~\cite{Mujal2021, Ghosh2021b}, a systematic characterization of their relative performance and resilience against noise remains largely unexplored.

In this work, we utilize a neutral atom array as a reservoir to investigate the impact of dynamical phases of matter, decoherence, and sampling noise on two QRC architectures: single-step-QRC (SS-QRC) and multi-step-QRC (MS-QRC). To address the measurement overhead problem, we incorporate the randomized measurement toolbox~\cite{Elben2023}, which allows for efficient measurements of multiple local observables. Our results demonstrate a striking contrast in robustness between the two architectures. Firstly, compared to MS-QRC, SS-QRC is less sensitive to dynamical phases of matter and decoherence noise due to its single-step nature. Additionally, in the presence of sampling noise, while the undermining of the convergence property in MS-QRC leads to a severe deterioration in its information processing capability, the SS-QRC framework remains remarkably robust. Our results suggest that the SS-QRC is a more suitable architecture for practical applications of the QRC formalism in the NISQ era.


\section{Physical model of the reservoir}

Figure~\ref{fig1}(a) illustrates the physical reservoir implemented using a neutral atom array trapped in optical tweezers~\cite{Kaufman2021, Manetsch2025, Browaeys2020}. The micrometer-scale spacing facilitates strong atom-atom interactions via global Rydberg excitation, while local light shifts enable both input encoding and flexible detection basis selection~\cite{Ma2025a}. Readout is performed via parallel fluorescence imaging on a single-photon CCD~\cite{DeOliveira2025, Anand2024a}. This platform is highly scalable, supporting thousands of atoms to access an exponentially large Hilbert space, and efficiently combines global control with precise local addressing through optical channel mapping~\cite{Bernien2017}.

In our model, quantum information is encoded in the ground state $\ket{g_i}$ and the Rydberg state $\ket{r_i}$ of the $i$-th atom. We define the computational basis as $\ket{0}_i = \ket{g_i}$ and $\ket{1}_i = \ket{r_i}$. The corresponding Hamiltonian of the atom array system is ($\hbar=1$)~\cite{Labuhn2016, Browaeys2020, Bernien2017}
\begin{equation}\label{eq:H}
    \hat{H} = \sum_{i=1}^N\Delta_i\hat{n}_i+\frac{\Omega}{2}\sum_{i=1}^N\hat{\sigma}^x_i+\sum_{i<j}\frac{V}{R_{ij}^6}\hat{n}_i\hat{n}_j,
\end{equation}
where $N$ denotes the total number of atoms, $\hat{n}_i=\dyad{r_i}$ is the population on the Rydberg state, $\Delta_i$ is the detuning with respect to the external laser driving, which induces the qubit state flip (Pauli $X$) $\hat{\sigma}^x_i=\dyad{r_i}{g_i}+\dyad{g_i}{r_i}$ with a  Rabi frequency $\Omega$, $V$ is the nearest-neighbor Rydberg blockade interaction strength, and $R_{ij}$ is the distance between the $i$-th and $j$-th atoms. 

For practical experiments on the atom array, decoherence is unavoidable. The dominant decoherence of the system originates from both the spontaneous decay of Rydberg states and the dephasing of the ground states, which can be described by the Lindblad jump operator~\cite{Bravo2022}
\begin{equation}
    \hat{L}_i = \sqrt{\gamma}\ket{g_i}(\alpha\bra{r_i}+\beta\bra{g_i})
\end{equation}
for the $i$-th atom. Here, $\gamma$ is the decay rate, and we set the coefficients $(\alpha,\beta)=(0.025,0.08)$. Then, the dynamics of the system is governed by the master equation
\begin{equation}\label{eq:master}
    \dot{\rho} = -\mathrm{i}[\hat{H}, \rho] + \sum_{i=1}^N\frac{1}{2}(2\hat{L}_i\rho \hat{L}_i^\dagger -\hat{L}_i^\dagger \hat{L}_i\rho-\rho \hat{L}_i^\dagger \hat{L}_i).
\end{equation}

Following the master equation, the physical reservoir functions as a non-linear feature map projecting inputs into a high-dimensional Hilbert space to enhance linear separability. Through sequential all-optical laser manipulation for input, the global Rydberg excitation, and readout, the generated features via the reservoir's evolution are then post-processed by classical computers. This approach offers a viable path toward demonstrating potential quantum advantages of NISQ devices without the overhead of optimizing system parameters. As illustrated by the quantum circuits in Figs.~\ref{fig1}(b) and (c), following the initialization of all qubits to $\ket{0}^{\otimes N}$ through optical pumping, quantum reservoir computing can be implemented through two different architectures, which we call SS-QRC and MS-QRC, respectively.

For the SS-QRC shown in Fig.~\ref{fig1}(b), the reservoir evolves for a single time step of duration $\tau$ for each input sample $\boldsymbol{x}^{(m)}$ from the training set $\{(\vb*{x}^{(m)}, y^{(m)})\}$. We assume a one-dimensional target $y^{(m)}$, though generalization to higher dimensions is straightforward. This evolution, driven by global Rydberg lasers is described by a CPTP map $\mathcal{E}(\boldsymbol{x}^{(m)})$. Subsequently, individual qubits are measured in arbitrary basis using addressed single-qubit rotations $R_i$, yielding the output vector $\boldsymbol{v}^{(m)}$. Finally, inference is performed on a classical computer using a linear layer, yielding the prediction $\tilde{y}^{(m)}=\boldsymbol{w}^\mathrm{T}\boldsymbol{v}^{(m)}$. This scheme is a quantum counterpart to classical extreme learning machines~\cite{Huang2015} and is frequently referred to as a quantum extreme learning machine~\cite{Mujal2021, Innocenti2023, Suprano2024}. Although the SS-QRC is inherently memoryless, incorporating a time-delayed embedding (sliding window) introduces a short-term memory determined by the window length~\cite{Butcher2013, Kornjaca2024}, allowing it to effectively model and analyze non-linear time-series data.

For the MS-QRC shown in Fig.~\ref{fig1}(c), the reservoir undergoes a multi-step evolution, each of duration $\tau$, to process a sequence of input samples $\boldsymbol{x}^{(1)},...,\boldsymbol{x}^{(m)}$. Each sample $\boldsymbol{x}^{(m)}$ is sequentially encoded into the system through a corresponding CPTP map $\mathcal{E}(\boldsymbol{x}^{(m)})$. After $m$ steps of evolution, the resulting state is measured to obtain the output vector $\boldsymbol{v}^{(m)}$, which is then fed to a classical computer for linear processing. Unlike the single-step approach, this sequential evolution endows the system with inherent short-term memory, rendering it naturally suitable for analyzing time-series data~\cite{Fujii2017}.

For both SS- and MS-QRC, each input $\vb*{x}^{(m)}$ is encoded into the detunings of the atoms by
\begin{equation}\label{eq:encoding}
    \Delta_i = \Delta + \eta x^{(m)}_{i}, \qquad i=1,\dots,N,
\end{equation}
where $\Delta$ represents the global bias detuning, $\eta$ is the input scaling parameter; $x^{(m)}_{i}$ is the $i$th element of the input $\vb*{x}^{(m)}$. To match the system size, input vectors with dimension smaller than $N$ are zero-padded (i.e., $x^{(m)}_i=0$ for $i > \dim(\vb*{x}^{(m)})$). The information is extracted from the reservoir by measuring the expectation values of single-site and two-site Pauli operators of $\sigma^x$ and $\sigma^z$ to form the output vector $\boldsymbol{v}^{(m)}$.

Both schemes are trained in the same way: the parameters of the reservoir remain fixed while only the weights of the final linear layer on the classical computer are optimized. By arranging the reservoir signals ${\boldsymbol{v}^{(1)},\dots, \boldsymbol{v}^{(m)}}$ into a matrix $\boldsymbol{V}=(\boldsymbol{v}^{(1)},\dots, \boldsymbol{v}^{(m)})^\mathrm{T}$ and the targets $y^{(1)},\dots,y^{(m)}$ into a vector $\boldsymbol{y}=(y^{(1)},\dots,y^{(m)})^\mathrm{T}$, the weights $\boldsymbol{w}$ of the output layer are usually trained using ridge regression \cite{Lukosevicius2012},
\begin{equation}
    \min_{\boldsymbol{w}}\ \|\boldsymbol{V}\boldsymbol{w}-\boldsymbol{y}\|_2^2+\lambda\|\boldsymbol{w}\|_2^2,
\end{equation}
where $\|\cdot\|_2$ denotes the $\ell_2$ norm, $\lambda$ controls the degree of regularization, and the ridge regression reduces to linear regression for $\lambda=0$.
The analytic solution is given by
\begin{equation}\label{eq:ridge}
    \boldsymbol{w} = (\boldsymbol{V}^\mathrm{T}\boldsymbol{V}+\lambda\boldsymbol{I})^{-1}\boldsymbol{V}^\mathrm{T}\boldsymbol{y},
\end{equation}
with $\boldsymbol{I}$ denoting the identity matrix. This solution is globally optimal, avoiding the issue of local minima from which traditional neural networks suffer. Furthermore, because the reservoir parameters remain fixed, QRC inherently bypasses the barren plateau problem. By restricting the learning process to a classical linear readout layer, the training phase becomes highly efficient, making QRC a particularly promising candidate for demonstrating quantum advantages in the NISQ era. Note that, unlike SS-QRC, during the training stage, a certain amount of initial time-series data is discarded in MS-QRC to wash out the influence of the initial conditions of the reservoir~\cite{Fujii2017}.

\section{Characterization of the reservoir}
\begin{figure}
    \centering
    \includegraphics[width=\linewidth]{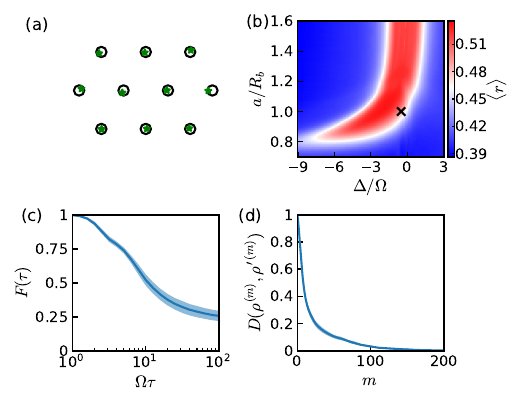}
    \caption{System configuration of an array of 10 atoms. (a) The triangular lattice of 10 atoms, where the positions of each atom (green stars) are normally distributed from the ideal lattice sites (empty circles). (b) The phase diagram of the atom array shown in (a), averaged over 1000 realizations of the atoms' positions at $\eta=0$. The parameters $(\Delta/\Omega=-0.5,\ a/R_b=1)$ marked by the black cross near the phase boundary are used for subsequent machine learning tasks. (c) Out-of-time-order correlator $F(\tau)$ as a function of duration $\tau$, with $\Delta/\Omega=-0.5$, $a/R_b=1$, $\eta=0$. (d) Convergence of MS-QRC characterized by the trace distance $D(\rho^{(m)},\rho'^{(m)})$, with the system starting from two different initial states $\rho^{(0)}$ and $\rho'^{(0)}$, and $\Delta/\Omega=-0.5$, $a/R_b=1$, $\gamma/\Omega=0.95$, $\tau=10/\Omega$, $\eta/\Omega=0.1$. The shaded area in (c-d) denotes the standard deviation across 10 realizations of the atoms' positions.}
    \label{fig2}
\end{figure}

Since the reservoir relies on the dynamical evolution to map the inputs to the high-dimensional space, we employ a triangular lattice configuration rather than a one-dimensional chain to facilitate rapid information scrambling across the array, as shown in Fig.~\ref{fig2}(a). Furthermore, breaking spatial symmetry is crucial for minimizing redundancy and accessing the full computational capacity of the Hilbert space. We leverage the positional disorder inherent in experimental setups to naturally break this symmetry. This is modeled by displacing atomic sites according to a normal distribution from the ideal lattice sites, with a standard deviation of $0.04a$, where $a$ is the lattice constant.

As the reservoir utilizes dynamical evolution for information processing and the degree of information scrambling is largely determined by the dynamical phases of matter~\cite{Lewis-Swan2019}, it is important to evaluate the impact of different dynamical phases of matter on the reservoir performance. The thermal phase scrambles local information rapidly to the whole system, while the information spreads much more slowly in the many-body localized phase, allowing the system to retain a memory of its initial conditions over long timescales. In previous studies~\cite{Martinez-Pena2021, Xia2022}, it has been shown that the performance of MS-QRC typically peaks at the dynamical phase transitions for time-series prediction due to the trade-off between nonlinear mapping and linear memory. Here, we characterize the dynamical phases of matter in a triangular array of $N=10$ atoms shown in Fig. \ref{fig2}(a) by the mean level spacing ratio $\ev{r}$ \cite{Oganesyan2007}. By exact diagonalization and arranging the eigenvalues $\{E_l\}$ of the Hamiltonian in Eq. \eqref{eq:H} in ascending order, one computes
\begin{align}
    \delta_l &= E_{l+1} - E_l, \\
    r_l &= \min\{\delta_l, \delta_{l+1}\}/\max\{\delta_l,\delta_{l+1}\},
\end{align}
and $\ev{r}$ is the mean value of $\{r_l\}$. For ergodic phase, $\ev{r}\approx 0.535$, and for localized phase, $\ev{r}\approx 0.386$ \cite{Atas2013}. The phase diagram of the atom array with respect to the lattice spacing $a$ in units of the Rydberg blockade radius $R_b$ and the global detuning $\Delta$ in units of the Rabi frequency $\Omega$, averaged over 1000 realizations of the atom positions at $\eta=0$ is shown in Fig.~\ref{fig2}(b), with the black cross near the dynamical phase transition marking the parameters $(\Delta/\Omega=-0.5,\ a/R_b=1)$ used in subsequent machine learning tasks. To ensure that information scrambling is dominated by the global bias detuning $\Delta$ rather than the input drive, we set the input scaling to $\eta/\Omega = 0.1$ in the following, unless otherwise specified.

In practice, the presence of decoherence imposes a trade-off on the interaction duration $\tau$. If $\tau$ is too short, the input information fails to scramble across the whole system; conversely, an excessively large $\tau$ leads to information loss via decoherence. To identify the optimal regime, we characterize the information spreading speed using the out-of-time-order correlator (OTOC)~\cite{Swingle2018, Kutvonen2020, Xia2022}. Although our reservoir is an open system, the closed-system OTOC provides a qualitative guidance for determining the appropriate $\tau$ required for sufficient scrambling before noise dominates the dynamics. The OTOC is defined as
\begin{equation}
F(\tau) = \frac{1}{N-1}\sum_{i=2}^{N}\ev{\hat{\sigma}^z_1(\tau)\hat{\sigma}^z_i(0)\hat{\sigma}^z_1(\tau)\hat{\sigma}^z_i(0)}_\beta,
\end{equation}
where $\hat{\sigma}_1^z(\tau)=e^{i\hat{H}\tau}\hat{\sigma}_1^z(0)e^{-i\hat{H}\tau}$, and $\ev{\cdot}_\beta$ denotes averaging over a thermal ensemble at temperature $1/\beta = k_\mathrm{B}T$. The OTOC $F(\tau)$ for the reservoir without input ($\eta=0$) at infinite temperature is shown in Fig. \ref{fig2}(c). It drops to $1/2$ at $\tau=10/\Omega$, indicating that local information has effectively scrambled throughout the system. Consequently, we select around $\tau=10/\Omega$ in the following as the evolution duration to balance effective scrambling against decoherence.

Additionally, the MS-QRC must possess the convergence property such that it can approximate time-invariant functions with fading memory~\cite{Chen2019c}. In open quantum systems, this convergence is physically guaranteed by the contractive nature of dissipative dynamics~\cite{Nielsen2010}. We characterize this convergence by evolving the system from two distinct initial states, $\rho^{(0)}=(\dyad{g})^{\otimes N}$ and a randomly selected mixed state $\rho'^{(0)}$, both subject to the same random input series sampled uniformly from $[-1, 1]$. The convergence of the resulting states after $m$ steps is quantified by the trace distance~\cite{Nielsen2010}
\begin{equation}
    D(\rho^{(m)},\rho'^{(m)}) = \frac{1}{2}\|\rho^{(m)}-\rho'^{(m)}\|_1,
\end{equation}
where $\|\cdot\|_1$ denotes the trace norm. As shown in Fig.~\ref{fig2}(d), the states converge after approximately 200 steps. Consequently, we discard the first 200 steps in subsequent simulations of MS-QRC to avoid the influence of initial conditions.

\begin{figure}
    \centering
    \includegraphics[width=\linewidth]{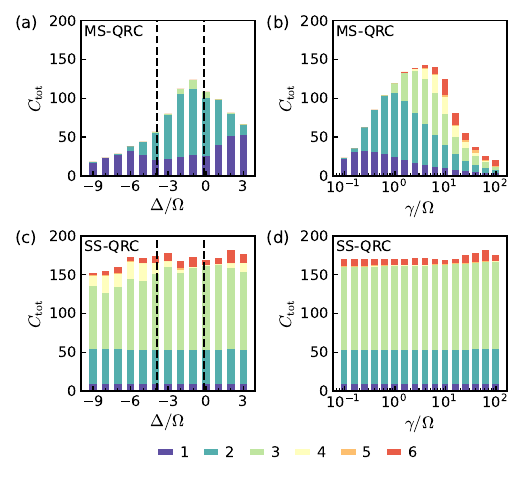}
    \caption{Performance with respect to the global bias detuning $\Delta$ and the decay rate $\gamma$. (a-b) The total information processing capacity (IPC), denoted as $C_\mathrm{tot}$, of MS-QRC with respect to the global bias detuning $\Delta$ and decay rate $\gamma$. $C_\mathrm{tot}$ is decomposed according to the order of nonlinearity, represented by bars with different colors. (c-d) The total IPC of SS-QRC with respect to the global bias detuning $\Delta$ and decay rate $\gamma$. The dashed lines in (a) and (c) indicates the phase boundaries. The results are averaged over 5 realization of the atoms' positions, with $a/R_b=1$, $\gamma/\Omega=0.95$, $\eta/\Omega=0.1$. The IPC are calculated using 1000 input time series uniformly sampled from [-1, 1] where the first 200 time steps are used for washout.}
    \label{fig3}
\end{figure}

To assess the general temporal-task-processing capabilities of SS-QRC and MS-QRC, providing a baseline before benchmarking specific applications, we calculate the information processing capacity (IPC)~\cite{Dambre2012, Kubota2021} of these two schemes. The IPC is a comprehensive metric for evaluating a reservoir's capability to transform the temporal input information across a basis of orthogonal functions. The total capacity, $C_\mathrm{tot} = \sum_{d=1}^\infty C_d$, is decomposed by the order of nonlinearity $d$, with $C_d$ representing contributions from $d$-th order polynomials. As shown in Fig.~\ref{fig3}(a), the MS-QRC capacity peaks near one of the dynamical phase transition regimes. Furthermore, Fig.~\ref{fig3}(b) reveals that the decay rate significantly impacts the performance of MS-QRC, which peaks around $\gamma/\Omega=4$. This can be explained by the convergence property of the reservoir: for a small decay rate, the reservoir converges too slowly for the MS-QRC to function, while for a large decay rate, the information dissipates into the environment. In contrast, since SS-QRC relies on single-step evolution rather than the multi-step evolution of MS-QRC, it exhibits remarkable resilience to both dynamical phases of matter and decoherence, as shown in Fig.~\ref{fig3}(c-d).

\section{QRC under sampling noise}
In experiments, expectation values are obtained through repeated state preparations and measurements. Given that each shot requires finite time, the total number of shots $N_s$ is necessarily limited, making sampling noise an unavoidable constraint. For practical implementations, characterizing QRC performance under these realistic conditions is essential. Furthermore, the QRC formalism requires measuring a large number of observables to extract sufficient information from the reservoir. To circumvent the prohibitive time cost of independent measurements, we leverage the randomized measurement toolbox~\cite{Elben2023}, which includes the randomized classical shadows~\cite{Huang2020} and derandomized classical shadows~\cite{Huang2021b}.

The randomized classical shadow formalism utilizes random Pauli measurements to efficiently estimate the expectation values of multiple local observables, where the number of shots $N_s$ required to predict the expectation values of $L$ observables within an additive error $\epsilon$ scales as $N_s \propto \log(L)3^w/\epsilon^2$, with $w$ denoting the weight of the observables~\cite{Huang2020}. The derandomized classical shadow protocol is a deterministic variant which replaces random basis selection with a greedy update procedure, ensuring performance that is at least as good as the randomized protocols and performs much better in some cases~\cite{Huang2021b}.

In this section, we assess the robustness of SS-QRC and MS-QRC under sampling noise, comparing the efficiency of the two measurement subroutines: randomized classical shadows versus derandomized classical shadows.

\subsection{Classification}
\begin{figure}
    \centering
    \includegraphics[width=\linewidth]{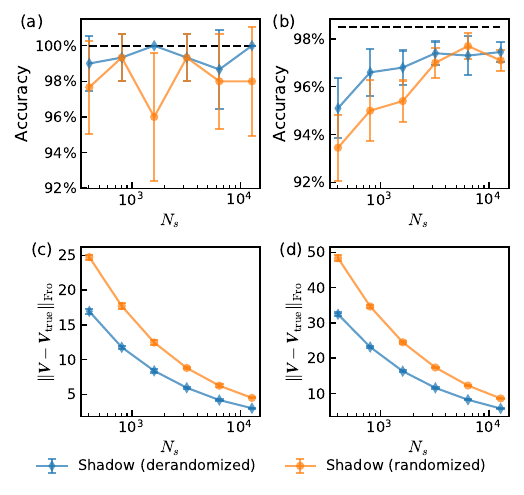}
    \caption{Performance of the SS-QRC on classification tasks, using the randomized measurement toolbox. (a) Classification accuracy on the Iris dataset, with $\Delta/\Omega=-0.5$, $a/R_b=1$, $\gamma/\Omega=0.95$, $\tau=10/\Omega$, $\eta/\Omega=0.1$. (b) Classification accuracy of SS-QRC on the entanglement-separability classification task, with $a/R_b=1$, $\gamma/\Omega=0.95$, $\tau=16/\Omega$. The dashed lines in (a-b) indicate the classification accuracy in the asymptotic limit ($N_s=\infty$). (c) Measurement error in the Iris dataset classification task, evaluated by the Frobenius norm $\|\boldsymbol{V}-\boldsymbol{V}_\mathrm{true}\|_\mathrm{Fro}$, where $\boldsymbol{V}_\mathrm{true}$ denotes the matrix of readout signals without sampling noise. (d) Measurement error in the entanglement-separability classification task. The error bars are standard deviations across 10 realizations of the measurement process.}
    \label{fig4}
\end{figure}

We first evaluate the impact of sampling noise on SS-QRC using two static benchmarks: the Iris dataset classification and the two-qubit entanglement-separability classification tasks.

The results of the Iris dataset classification are shown in Fig.~\ref{fig4}(a). In the asymptotic limit ($N_s = \infty$), the SS-QRC achieves 100\% classification accuracy. For finite $N_s$, the classification accuracy  maintains over 98\% for the derandomized classical shadow protocol, consistently outperforming the randomized scheme on average. This superiority is confirmed by the lower measurement error shown in Fig.~\ref{fig4}(c).


To test the system on a more challenging task, we consider the separability-entanglement classification of two-qubit states. Similar to Ref. \cite{Lu2018a}, we randomly generate 1000 two-qubit states $\rho^{(m)}$ and prepare the dataset as $\{(\rho^{(m)}, \alpha^{(m)}; y^{(m)})\}_{m=1}^{1000}$, where $y^{(m)}$ is the binary label of the $m$-th two-qubit state $\rho^{(m)}$ obtained by the positive partial transpose criterion~\cite{Horodecki2009}, and $\alpha^{(m)}$ is a quantity denoting the geometric information of the state $\rho^{(m)}$ extracted from a convex hull approximation~\cite{Lu2018a}.


In our simulation, we prepare the initial state of the first two Rydberg atoms directly as $\rho^{(m)}$ and that of the remaining 8 atoms as a product state of $\dyad{g_i},\ i=2,\dots,8$. The geometric information $\alpha^{(m)}$ is encoded into the detuning of the 10 atoms by $\Delta_i=\alpha^{(m)}$ for each $i$. The readout signals are obtained by measuring the single-site and two-site Pauli operators $\sigma^x$ and $\sigma^z$ of the remaining 8 atoms. 

As shown in Fig. \ref{fig4}(b), the classification accuracy in the asymptotic limit ($N_s=\infty$) is 98.5\%, and the accuracy under sampling noise does not degrade too much for $N_s=1\times 10^4$. For a small number of measurement shots, the classification accuracy based on the derandomized classical shadow outperforms the randomized protocol on average; while they are comparable when $N_s$ is large. As seen in Fig.~\ref{fig4}(d), the derandomized classical shadow provides a significant reduction in measurement error compared to the randomized shadow when $N_s$ is small.

\subsection{Time-series prediction}
\begin{figure}
    \centering
    \includegraphics[width=\linewidth]{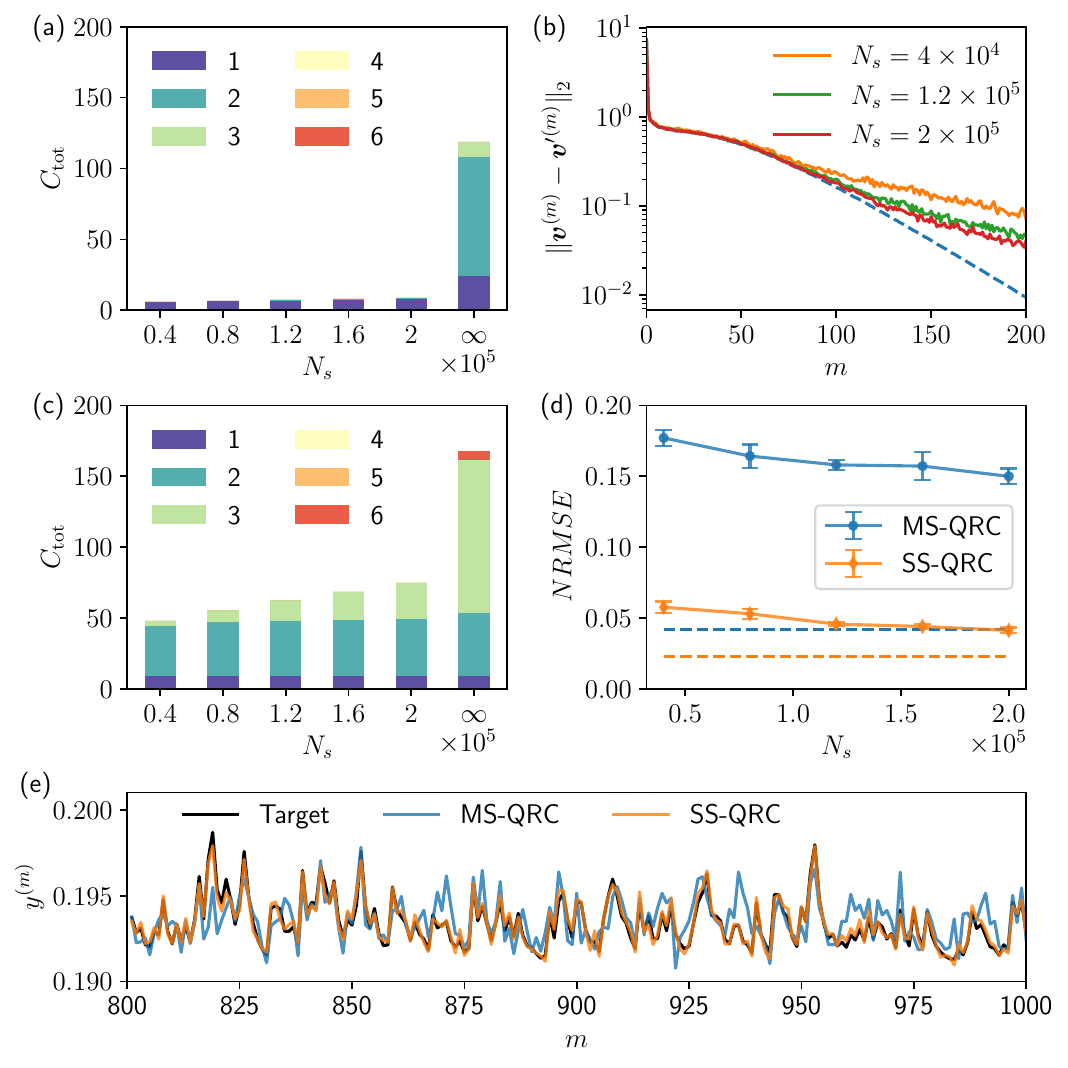}
    \caption{Performance of MS-QRC and SS-QRC on time-series prediction tasks under sampling noise using the derandomized classical shadow protocol. (a) The IPC of MS-QRC versus the number of measurement shots $N_s$. (b) Convergence of MS-QRC characterized by $\|\boldsymbol{v}^{(m)}-\boldsymbol{v}'^{(m)}\|_2$, with the reservoir starting from two different initial states. The dashed blue curve corresponds to the asymptotic limit. (c) The IPC of the SS-QRC. The results in (a) and (c) are averaged over 5 realizations of the atoms positions. (d) The normalized root mean square error (NRMSE) of MS-QRC and SS-QRC on the second-order NARMA task, with the dashed lines corresponding to $N_s=\infty$ for the two models. Error bars are standard deviations across 10 realizations of the derandomized classical shadow protocol. (e) The predictions of MS-QRC and SS-QRC on the second-order NARMA task at $N_s=2\times 10^5$. The parameters used in (a-e) are $\Delta/\Omega=-0.5$, $a/R_b=1$, $\gamma/\Omega=0.95$, $\eta/\Omega=0.1$.}
    \label{fig5}
\end{figure}

Since the derandomized classical shadow outperforms the randomized protocol, we adopt it as the measurement subroutine in the following and evaluate the temporal processing capability of MS-QRC and SS-QRC under sampling noise. We first evaluate the IPC of these two QRC architectures. As shown in Fig.~\ref{fig5}(a), while MS-QRC exhibits high IPC in the asymptotic limit ($N_s=\infty$), sampling noise drastically reduces its capacity. In particular, its nonlinear processing capability nearly vanishes in the presence of sampling noise. This can be explained by the undermining of the convergence property of the MS-QRC under sampling noise, as shown in Fig.~\ref{fig5}(b). Note that sampling noise affects the reservoir's output signals $\boldsymbol{v}^{(m)}$ used for learning, not the quantum CPTP map itself. In contrast, since the SS-QRC has no convergence requirements, its IPC maintains up to the third order under finite $N_s$, as shown in Fig.~\ref{fig5}(c).

The robustness of SS-QRC against sampling noise is further demonstrated using a specific time-series prediction task, the second-order nonlinear autoregressive moving average (NARMA) task~\cite{Fujii2017},
\begin{equation}
\begin{aligned}
    y^{(m+1)} =&\ 0.4y^{(m)} + 0.4y^{(m)}y^{(m-1)} \\
    &+ 0.6(u^{(m)})^3 + 0.1,
\end{aligned}
\end{equation}
where the input $u^{(m)}$ is uniformly sampled from $[0, 0.2]$. With a total of 1000 time steps, we use the first 200 steps for washout, the intermediate 600 steps for training, and the last 200 steps for testing. The performance on the second-order NARMA task is evaluated by the normalized root mean square error (NRMSE), defined as
\begin{equation}
    NRMSE = \frac{1}{A}\sqrt{\frac{1}{200}\sum_{m=801}^{1000}(\tilde{y}^{(m)}-y^{(m)})^2},
\end{equation}
where the normalization parameter is defined as $A=\max\{y^{(m)}\}-\min\{y^{(m)}\}$. Compared to SS-QRC, the NRMSE of MS-QRC increases more significantly under sampling noise, as shown in Fig.~\ref{fig5}(d). The predicted time series at $N_s = 2 \times 10^5$ are shown in Fig.~\ref{fig5}(e). While the predictions of MS-QRC deviate significantly from the target series, the SS-QRC predictions match the target series closely. These results suggest that the SS-QRC architecture is more robust against sampling noise and is a preferred candidate for near-term applications.


\section{Discussion and Conclusion}
While Rydberg atom arrays have been demonstrated as a viable platform for QRC~\cite{Bravo2022, Kornjaca2024}, a systematic characterization of how realistic constraints in the Rydberg platform impact the performance of different QRC architectures, specifically the MS-QRC and SS-QRC, remains to be established. In this work, we have provided a comprehensive comparative analysis of MS-QRC and SS-QRC with a Rydberg atom array implementation, focusing on the dynamical phases of matter, decoherence, and sampling noise.

Our findings reveal a significant distinction in the robustness of these two QRC architectures. We demonstrate that whereas MS-QRC is highly sensitive to the underlying dynamical phases of matter and decoherence, SS-QRC exhibits remarkable resilience to these system configurations. Furthermore, we utilize the derandomized classical shadow as a subroutine to mitigate the measurement overhead of the reservoir. In the presence of sampling noise, the convergence property of MS-QRC is undermined, leading to a significant decrease in its IPC. Conversely, the SS-QRC still largely maintains its IPC. These results suggest that while both models exhibit comparable performance in the asymptotic limit ($N_s = \infty$), the SS-QRC is more robust against sampling noise and is thus more suitable for NISQ-era applications.

Future improvements could involve the application of advanced denoising techniques, such as singular value decomposition, specialized signal filtering~\cite{Ahmed2025}, or eigentask techniques~\cite{Hu2023}, to further mitigate the effects of statistical fluctuations.

The source code for this study is available in the GitHub repository~\cite{Liu2024b}.

\begin{acknowledgments}
This work was funded by the National Key R\&D Program (Grant Nos.~2021YFA1402004 and 2022YFA1204704), the National Natural Science Foundation of China (Grants No. 92465201, T2325022, U23A2074 and 92265108), Quantum Science and Technology-National Science and Technology Major Project (Grants 2021ZD0303200, 2021ZD0301500), and the CAS Project for Young Scientists in Basic Research (No. YSBR-049). This work was also supported by the Fundamental Research Funds for the Central Universities and USTC Research Funds of the Double First-Class Initiative. The numerical calculations in this paper were performed at the Supercomputing Center of USTC. This work was partially carried out at the USTC Center for Micro and Nanoscale Research and Fabrication.
\end{acknowledgments}

\appendix
\section{Details of data encoding}
For general non-temporal datasets $\{(\boldsymbol{x}^{(m)}, y^{(m)})\}$, the input vectors $\boldsymbol{x}^{(m)}$ are directly mapped onto the atomic array via Eq.~\eqref{eq:encoding}. In contrast, for one-dimensional time-series tasks $\{(u^{(m)}, y^{(m)})\}$, the scalar inputs must first be transformed into an $N$-dimensional vector $\boldsymbol{x}^{(m)}$ before encoding. The construction of $\boldsymbol{x}^{(m)}$ differs between the two architectures. For SS-QRC, the temporal sequence is embedded using a sliding window of length $N$, where $N$ denotes the number of atoms, such that $\boldsymbol{x}^{(m)} = (u^{(m-N+1)}, \dots, u^{(m-1)}, u^{(m)})^\mathrm{T}$. Conversely, for MS-QRC, the reservoir is driven by a uniform broadcasting of the instantaneous input, such that $\boldsymbol{x}^{(m)}$ is an $N$-dimensional vector with all elements equal to $u^{(m)}$. Following this transformation, the resulting vectors $\boldsymbol{x}^{(m)}$ for both architectures are encoded into the reservoir according to Eq.~\eqref{eq:encoding}.


%

\end{document}